\documentstyle[epsf,rotate]{mn}
\newif\ifAMStwofonts
\AMStwofontstrue

\def\O{$\Omega$}
\input wasyfont
 
\def\ar{{$\rightarrow$}}

\def\D{{$\Delta$ }} 
\def\Dd{{$\Delta$}}          

\ifoldfss
  \ifCUPmtlplainloaded \else
    \NewTextAlphabet{textbfit} {cmbxti10} {}
    \NewTextAlphabet{textbfss} {cmssbx10} {}
    \NewMathAlphabet{mathbfit} {cmbxti10} {} 
    \NewMathAlphabet{mathbfss} {cmssbx10} {} 
  \fi
  \ifAMStwofonts
    \ifCUPmtlplainloaded \else
      \NewSymbolFont{upmath} {eurm10}
      \NewSymbolFont{AMSa} {msam10}
      \NewMathSymbol{\upi}     {0}{upmath}{19}
      \NewMathSymbol{\umu}     {0}{upmath}{16}
      \NewMathSymbol{\upartial}{0}{upmath}{40}
      \NewMathSymbol{\leqslant}{3}{AMSa}{36}
      \NewMathSymbol{\geqslant}{3}{AMSa}{3E}

    \fi
  \fi
\fi 

\ifnfssone
  \newmathalphabet{\mathit}
  \addtoversion{normal}{\mathit}{cmr}{m}{it}
  \addtoversion{bold}{\mathit}{cmr}{bx}{it}
  \newmathalphabet{\mathbfit} 
  \addtoversion{normal}{\mathbfit}{cmr}{bx}{it}
  \addtoversion{bold}{\mathbfit}{cmr}{bx}{it}
  \newmathalphabet{\mathbfss} 
  \addtoversion{normal}{\mathbfss}{cmss}{bx}{n}
  \addtoversion{bold}{\mathbfss}{cmss}{bx}{n}
  \ifAMStwofonts
    \ifCUPmtlplainloaded \else
      \UseAMStwoboldmath
      \makeatletter
      \new@mathgroup\upmath@group
      \define@mathgroup\mv@normal\upmath@group{eur}{m}{n}
      \define@mathgroup\mv@bold\upmath@group{eur}{b}{n}
      \edef\UPM{\hexnumber\upmath@group}
      \new@mathgroup\amsa@group
      \define@mathgroup\mv@normal\amsa@group{msa}{m}{n}
      \define@mathgroup\mv@bold\amsa@group{msa}{m}{n}
      \edef\AMSa{\hexnumber\amsa@group}
      \makeatother
      \mathchardef\upi="0\UPM19
      \mathchardef\umu="0\UPM16
      \mathchardef\upartial="0\UPM40
      \mathchardef\leqslant="3\AMSa36
      \mathchardef\geqslant="3\AMSa3E
    \fi
  \fi
\fi 

\ifnfsstwo
  \DeclareMathAlphabet{\mathbfit}{OT1}{cmr}{bx}{it}
  \SetMathAlphabet\mathbfit{bold}{OT1}{cmr}{bx}{it}
  \DeclareMathAlphabet{\mathbfss}{OT1}{cmss}{bx}{n}
  \SetMathAlphabet\mathbfss{bold}{OT1}{cmss}{bx}{n}
  \ifAMStwofonts
    \ifCUPmtlplainloaded \else
      \DeclareSymbolFont{UPM}{U}{eur}{m}{n}
      \SetSymbolFont{UPM}{bold}{U}{eur}{b}{n}
      \DeclareSymbolFont{AMSa}{U}{msa}{m}{n}
      \DeclareMathSymbol{\upi}{0}{UPM}{"19}
      \DeclareMathSymbol{\umu}{0}{UPM}{"16}
      \DeclareMathSymbol{\upartial}{0}{UPM}{"40}
      \DeclareMathSymbol{\leqslant}{3}{AMSa}{"36}
      \DeclareMathSymbol{\geqslant}{3}{AMSa}{"3E}
    \fi
  \fi
\fi 

\ifCUPmtlplainloaded \else
  \ifAMStwofonts \else 
    \def\upi{\pi}
    \def\umu{\mu}
    \def\upartial{\partial}
  \fi
\fi

\title{Orbital dynamics of three-dimensional bars: \\II. Investigation of the
  parameter space}
\author[Ch.~Skokos et al.]
{Ch.~Skokos,$^1$ P.A.~Patsis,$^1$ E.~Athanassoula$^2$\\
$^1$Research Center of Astronomy, Academy of Athens, Anagnostopoulou 14,
  GR-10673 Athens, Greece\\
$^2$Observatoire de Marseille, 2 Place Le Verrier, F-13248 Marseille Cedex 4,
  France}
\date{Accepted ....
      Received ....;
      in original form ....}
\pagerange{\pageref{firstpage}--\pageref{lastpage}}
\pubyear{2001}

\begin{document}

\maketitle

\label{firstpage}

\begin{abstract}
  We investigate the orbital structure in a class of 3D models of
  barred galaxies.  We consider different values of the pattern speed,
  of the strength of the bar and of the parameters of the central
  bulge of the galactic model.  The morphology of the stable orbits in
  the bar region is associated with the degree of folding of the
  x1-characteristic. This folding is larger for lower values of the
  pattern speed. The elongation of rectangular-like orbits belonging
  to x1 and to x1-originated families depends mainly on the pattern
  speed.  The detailed investigation of the trees of bifurcating
  families in the various models shows that major building blocks of
  3D bars can be supplied by families initially introduced as unstable
  in the system, but becoming stable at another energy
interval. In some models without radial and vertical 2:1 resonances
  we find, except for the x1 and x1-originated families, also families
  related to the z-axis orbits, which support the bar.  Bifurcations
  of the x2 family can build a secondary 3D bar along the minor axis
  of the main bar.  This is favoured in the slow rotating bar case.
\end{abstract}


\section{Introduction}
Barred galaxies have bars of very different strength, ranging from the
weak bars of SAB galaxies to the strong bars of e.g. NGC~1365
\cite{lpo99}. They may
have large, small, or no bulge components in their
centers. The possibility that bars in late-type barred spiral galaxies
end at their inner Lindblad resonance (hearafter ILR) 
has also been considered (Lynden-Bell 1979, Combes \&
Elmegreen 1993, Polyachenko \& Polyachenko 1994)
and this would imply that in some cases bars may have corotation far
beyond their ends. It is thus important to understand whether
and to what extent the orbital structure changes with the
basic parameters in the models. We investigate this here
using a class of models, the individual representatives of which
differ in their central mass concentration and in the pattern speed
and strength of the bar. 

We follow the evolution of  all the families of
periodic orbits we think may play a role in the dynamics and morphology of
bars and peanuts. We believe we indeed have all the 
main families for two reasons. First, the edge-on profiles
of the galaxies are mainly affected by the vertical bifurcations up to the 4:1
vertical resonance. Beyond this resonance the orbits of 
the bifurcating 3D families remain
close to the equatorial plane and thus do not characterize the edge-on
morphology. Second, families bifurcated at the n:1 radial resonances
for $n>4$ do not in general support the bar (e.g. Contopoulos \& Grosb{\o}l
1989, Athanassoula 1992).

The models presented here are static, but they may be
viewed as corresponding to individual phases of an evolutionary
process of the dynamical evolution of a galaxy within a Hubble
time. Therefore, a complete investigation of the dynamical system is
necessary in order to find all orbits possibly associated with the
presence of specific morphological features.

In the first paper of this series (Skokos, Patsis \& Athanassoula 2002,
hereafter paper I) we presented the basic families in a model composed
of a Miyamoto disc of length scales A=3 and B=1, a Plummer sphere 
of scale length 0.4 for a
bulge and a Ferrers bar of index 2 and axial ratio $a:b:c =
6:1.5:0.6$.  The masses of the three components satisfy
\( G(M_{D}+M_{S}+M_{B})=1 \) and are given in Table~\ref{tab:models}.
 The length unit is 1~kpc, the time unit
1~Myr and the mass unit $ 2\times 10^{11} M_{\odot}$.

In the present paper we compare the orbital structure of our basic model
with those
encountered in five more models. Our models, including the fiducial
model A1 of paper I, are described in Table~\ref{tab:models}. G is the
gravitational constant, M$_D$, M$_B$, M$_S$ are the masses of the
disk, the bar and the bulge respectively, $\epsilon_s$ is the scale
length of the bulge, \O$_{b}$ is the pattern speed of the bar,
$E_j$(r-IILR) and $E_j$(v-ILR) are the values of the Jacobian for the
inner radial ILR and the vertical 2:1 resonance respectively,
and $R_c$ is the corotation radius.
\begin{table*}
\caption[]{Parameters of our models. G is the gravitational constant,
M$_D$, M$_B$, M$_S$ are the masses of the disk, the bar and the bulge
respectively, $\epsilon_s$ is the scale length of the bulge, \O$_{b}$
is the pattern speed of the bar, $E_j$(r-IILR) is the Jacobian for the
inner radial ILR, $E_j$(v-ILR) is the Jacobian for the vertical 2:1
resonance, $R_c$ is the corotation radius. The comment in the last
column characterizes the model  in order to facilitate
its identification. }
\label{tab:models}
\begin{flushleft}
\begin{tabular}{cccccccccc}
model name& GM$_D$ & GM$_B$ & GM$_S$ & $\epsilon_s$ & \O$_{b}$ & $E_j$(r-IILR)
&$E_j$(v-ILR)  
& $R_c$ & comments\\ 
\hline
A1 & 0.82 &  0.1  & 0.08 & 0.4 &  0.0540 & -0.441 & -0.360&    6.13 &
fiducial \\
A2 & 0.82 &  0.1  & 0.08 & 0.4 &  0.0200 & -0.470 & -0.357&   13.24 &
slow bar \\
A3 & 0.82 &  0.1  & 0.08 & 0.4 &  0.0837 & -0.390 & -0.364&    4.19 &
fast bar \\
B  & 0.90 &  0.1  & 0.00 & --  &  0.0540 &  --    &  --   &    6.00 &
no bulge \\
C  & 0.82 &  0.1  & 0.08 & 1.0 &  0.0540 &  --    & -0.364&    6.12 &
extended bulge \\
D  & 0.72 &  0.2  & 0.08 & 0.4 &  0.0540 & -0.467 & -0.440&    6.31 &
strong bar \\
\hline
\end{tabular}
\end{flushleft}
\end{table*}

This paper is organized as follows: In section 2 we discuss models with fast,
or with slow bars. Section 3 introduces a model with no 2:1 resonances, 
section 4 a model with vertical but no radial ILR, and section 5 a model with 
a massive
bar. We conclude in section 6.

\section{The effect of pattern speed}
\subsection{A slow rotating bar}
Model A2 is the same as model A1 in everything, except for the pattern
speed, \O$_b=0.02$, which is less than half that of model A1. The
corotation in this model is at 13.24, rather than at 6.13 as in model A1,
and the outer inner Lindblad resonance (OILR) is now at 6.1,
i.e. roughly the end of the bar or the corotation distance of the
models with \O$_{b}$ = 0.054.

\begin{figure}
\epsfxsize=9.0cm \epsfbox{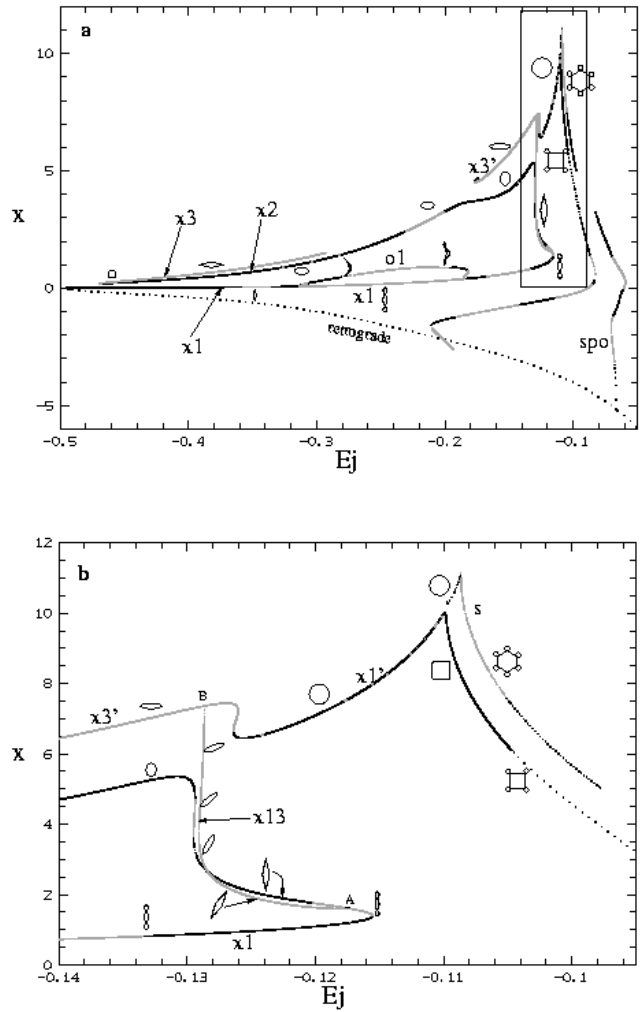}
\caption[]{Characteristic diagram for the 2D families of model
  A2. Grey parts of the lines show the unstable parts of the
  families. In (b) we give an enlargement of the area included in the
  frame in (a).}
\label{a2-x1char}
\end{figure}

The changes in the dynamical behaviour are much more important than a
stretching of the corotation radius by a factor larger than two and an
enlargement of the x2-x3 loop in the characteristic and in the
stability curves of the model. New bifurcations and new gaps are
introduced, while the morphology of some of the existing families
changes drastically.  The differences are so big as to introduce
nomenclature issues. Let us start our examination of the main simple-periodic
families 
and of their bifurcations with the help of the characteristic diagram
for planar orbits, shown in Fig.~\ref{a2-x1char}.
Following the convention introduced in paper I, we draw with a black
line the parts of the characteristics which correspond to stable parts
of the families, while grey is associated with instability. There are
two main characteristics, or rather families of characteristics. The
lower one is confined to the region below $x \approx 5.5$. It is divided from
the upper characteristic by  
a gap, occurring roughly at $E_j = -0.128$. There are are also a number
of 3D families bifurcating from these characteristics, of which the
most important ones will be described at the end of this section.

The main feature of the characteristic diagram is a continuous curve
constituted by the simple-periodic 2D families x1, x2 and x3. We will follow
it counter-clockwise. Starting 
close to $E_j=-0.5$ for $x=0$ we walk along the characteristic of the 
typical x1 family. The
orbits there are elliptical-like and support the bar.

At the first S \ar U transition of x1 the family x1v1 is
bifurcated. That means we have reached at this energy the vertical 2:1
resonance. It has a similar evolution as in the fiducial case (paper
I), but it is complex unstable for a considerably smaller energy
range. This affects strongly the vertical profile of the model
(Patsis, Skokos \& Athanassoula, 2002a hereafter paper III). Since it
is a 3D family it is not included in Fig.~\ref{a2-x1char}.

The first radial bifurcation occurs at $E_j \approx -0.31$ and gives
the family o1. This is stable for a tiny $E_j$ interval, just after
the S\ar U transition. It then follows a S\ar U\ar S\ar U sequence and
ends again on x1.  Thus this family builds a bubble, 
both in the characteristic and the 
stability diagram, together with x1 or with its indices, as did family 
  t1 in model
A1.  Its morphology, however, shows that it is related to a radial 1:1
resonance \cite{pape83}, since both cuts with the $y=0$ axis are for
$x>0$ (alternatively $x<0$), so that it can be viewed as a distorted
circle. It nevertheless has three tips or `corners', of which the two
close to the $y$ axis are very sharp and for large $E_j$ values they
develop loops.  This morphological evolution is reflected by the small
orbits drawn close to the o1 curve on Fig.~\ref{a2-x1char}a. 

The next
S\ar U transition brings in the system x1v3. This is a 3D family, so
again it is not included in the diagram. When x1 becomes again stable
close to $E_j \approx -0.134$, its orbits have developed loops along
the major axes of the ellipses. Since we are already at the area
included in the frame in Fig.~\ref{a2-x1char}a, it is easier to follow
the evolution of the families on the characteristic in
Fig.~\ref{a2-x1char}b. We observe that close to $E_j =-0.115$ the
curve has a bend and continues towards lower $E_j$ and higher $x$
values.  On the bend x1 orbits are still very elongated with
loops at the $y$ axis, as noted by a x1 orbit drawn there. The x1 family
has developed these loops well before the bend. Between $E_j=-0.115$
and $E_j=-0.13$, at the rising part of the characteristic, towards
lower $E_j$ values, the x1 orbits become again
ellipse-like and their loops vanish  
(for the time being we forget about the gray branch we
observe at the same area).  Meanwhile, the characteristic curve has
two more bends at $x$ roughly 3 and 5 respectively and for almost
the same value of $E_j \approx -0.13$, and then follows the long
branch towards lower $E_j$ values, which reaches $E_j \approx -0.47$.
On this branch and close to $E_j =-0.13$ the x1 orbits have small
ellipticities and become even rounder as we move to lower $E_j$
values. Finally after $E_j \approx -0.17$ the orbits are elongated
{\em along the minor axis} of the bar, and are stable (except for
$-0.23 < E_j < -0.2$), i.e.  they belong to the x2 family.  At $E_j
\approx -0.47$ the curve folds again and continues its journey towards
larger $E_j$ values. The orbits at this branch are typical x3 orbits
and exist until $E_j \approx -0.29$, where they change multiplicity. Thus in
this model the x2 and 
the x3 families are {\em continuations} of the x1, the transition
being made by circular and circular-like orbits, rather than by a gap
as in the standard cases (Contopoulos \& Grosb{\o}l 1989, Athanassoula
1992a, paper I etc.).

At $E_j \approx -0.23$  the stability index
associated with the 3D bifurcations intersects the $-2$ axis. So we have the
bifurcation of a new family with the {\em same multiplicity}. We call this 
family x2v1. We emphasize the fact that this is a simple periodic family,
since in model A1 (paper I) we had already encountered a 3D bifurcation of x2
(family x2mul2), which, however, is of multiplicity 2. Since this new family
is a direct bifurcation of x2 at the S\ar U transition close to $E_j = -0.23$,
as we move towards larger values of $E_j$, it inherits the stability of the
parent family, i.e.  it is stable.
\begin{figure}
\vspace{0cm}
\epsfxsize=8.0cm \epsfbox{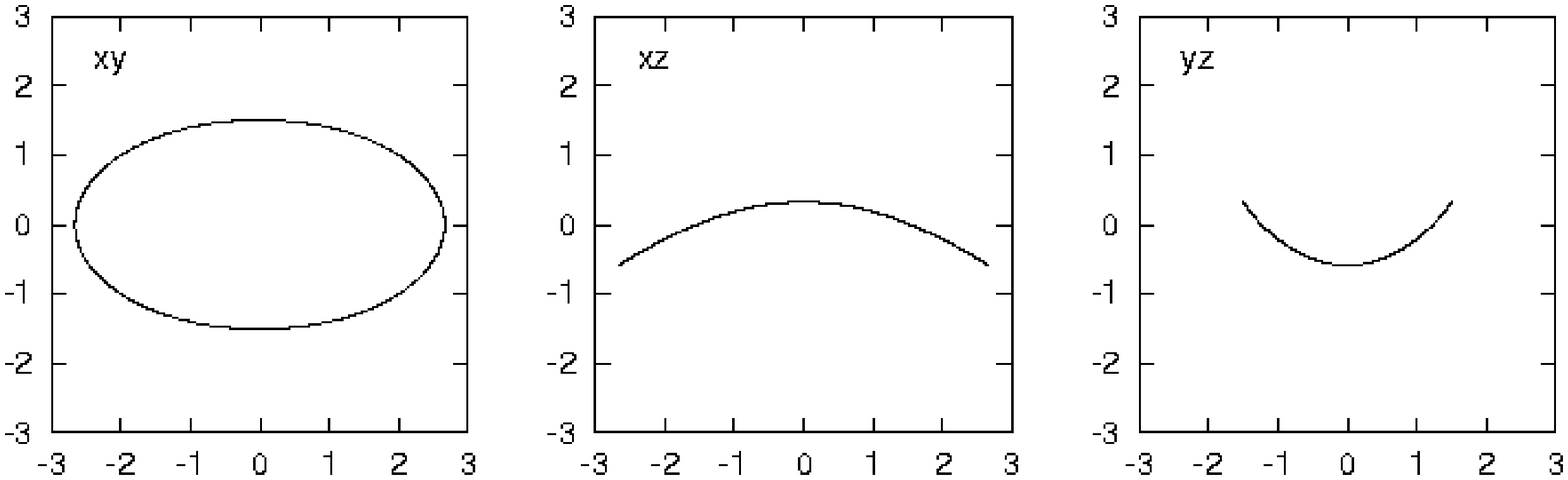}
\vspace{0cm}
\caption[]{Stable orbit of the x2v1 family. } 
\label{o-x2v1}
\end{figure}
It stays stable for a large energy interval, $-0.23<E_j<-0.18$, which
means that it is a family that can affect the morphology of the
galaxy. Its morphology can be seen in Fig.~\ref{o-x2v1}. As we can see
this family can support a peanut-like feature, which, however, is
elongated not along the major but along the {\em minor} axis of the
main bar. If such orbits are populated in a real galaxy, then they
will support a 3D stellar inner bar with a `x2 orientation'.

\begin{figure}
\rotate[r]{ 
\epsfxsize=4.0cm \epsfbox{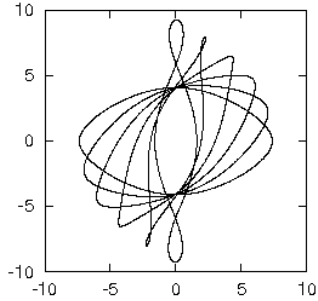}
}
\caption[]{Successive x13 orbits. They are all unstable. }
\label{o-x13}
\end{figure}

Close to the part of the x1 characteristic for $-0.13<E_j<-0.115$, where the
curve folds and extends towards lower energies (Fig.~\ref{a2-x1char}b), we
have, besides the `x1 part', a gray branch (unstable orbits) that bridges the
main loop with another branch of the characteristic diagram existing at the
same energies and for larger $x$ values. If this bridge was missing then we
would have a classical type 2 gap as at the radial 4:1 resonance regions
(Contopoulos \& Grosb{\o}l 1989). What we have now could be called a
pseudo-gap. The orbits of this branch are unstable and belong to a
family we call x13, since it starts for low $x$ values as x1 at point
`A' (Fig.~\ref{a2-x1char}b) and reaches at `B' a horizontal branch,
which is the characteristic curve of a x3-like family.   x13 is a radial bifurcation
in $\dot{x}$, so the curve we give in Fig.~\ref{a2-x1char} for this
family is just the projection of its characteristic in the ($E_j$,$x$)
plane. The morphology of these orbits is expected to be related with
inclined ellipses, whose major axis shifts from being parallel to the
bar major axis (for members on or near the major loop characteristic)
to parallel to the bar minor axis (for members on or near the
x3$^{\prime}$ characteristic).  The shift happens in a small energy
interval, in which the x1 orbits have the longest projections on the
$y$ axis.  Successive orbits of x13, as we move from `A' to `B'
(Fig.~\ref{a2-x1char}b), are given in Fig.~\ref{o-x13}. 
The evolution of the stability indices
of x1 in this area follow every possible complication one could
imagine in order to avoid bifurcating a stable family with similar
morphology. Due to this `conspiracy' it was not possible for us to
find a stable  x13-like family.

\begin{figure}
\rotate[r]{       
\epsfxsize=5.2cm \epsfbox{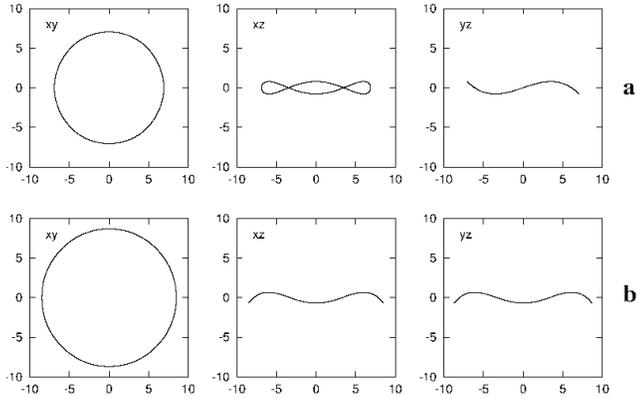}
}
\caption[]{Stable orbits of the families x1$^{\prime}$v4 (a), and
x1$^{\prime}$v5 (b). }
\label{A2-3dbifs}
\end{figure}

The second part of the characteristic diagram, at the same energies as the 
`x1 part' and for higher $x$ values, has orbits which are x3-like.
These orbits are ellipses elongated along the minor axis of the bar
and are almost everywhere unstable, except for a tiny part of the
characteristic for $E_j \approx -0.175$. We thus called them x3$^{\prime}$.
Moving along the x3$^{\prime}$ branch of the characteristic towards
larger $E_j$ values, we encounter a step-like feature in the curve
(Fig.~\ref{a2-x1char}b) and beyond it we have planar orbits, which can
be easily described as prograde quasi-{\em circular} orbits. Their
general dynamical properties and their relation with other families at
the area resemble those of the x1 family.  So this family is a kind of
continuation of x1, which we call x1$^{\prime}$ (as we called, for
lower energies, the continuation of the x3 family x3$^{\prime}$). The
stability indices of x1$^{\prime}$ oscillate and at the tangencies
with the $b=-2$ axis the 3D families x1$^{\prime}$v4, x1$^{\prime}$v5
etc. are born. We call them like this because their morphology on the
$(x,z)$ and $(y,z)$ projections resembles the morphology of the x1v4
and x1v5 families of the fiducial case.  The bifurcated 3D families
remain as stable close to the equatorial plane, i.e. they do not
characterize the vertical profile of the model, although they have
large stable parts. It is important to note that in this case the
shape of the x1$^{\prime}$ orbits -- and of the $(x,y)$ projections of
x1$^{\prime}$v4 and x1$^{\prime}$v5 as well -- are {\em not} elongated
along the major axis of the bar, but are quasi-circular. Thus, they
{\em do not} enhance the bar towards the corotation radius
(13~kpc). This can be seen in Fig.~\ref{A2-3dbifs}.

The characteristic of x1$^{\prime}$, as in the case of model A1 for
x1, has a local maximum at $E_j \approx -0.11$. At the decreasing
branch (lower $x$'s for larger $E_j$ values) the orbits of the family
develop `corners'. The usually rectangular-like orbits found in the
4:1 region (cf. Fig.~2g in paper I) are for this model
square-like. x1$^{\prime}$ has a stable part just after the turning
point, while in model A1 the decreasing branch is almost everywhere
unstable. For yet larger energies, when the orbits at their four
apocentra have loops, x1$^{\prime}$ is unstable.

\begin{figure}
\rotate[r]{   
\epsfxsize=6.0cm \epsfbox{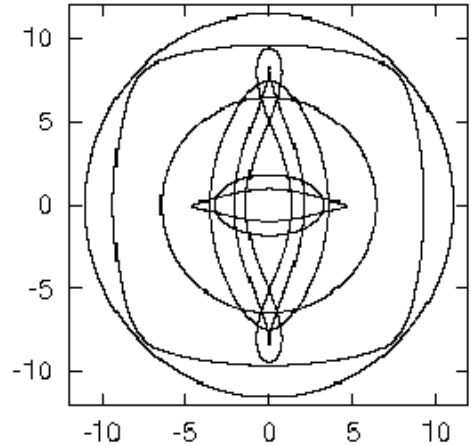}
}
\caption[]{Stable orbits for model A2. }
\label{a3model}
\end{figure}

As can be seen in Fig.~\ref{a2-x1char}b the gap at $E_j \approx -0.11$
is a real type 2 gap, the upper branch of which has stable circular
orbits at the `increasing $x$' part and unstable hexagonal orbits at the
decreasing part following it at larger $E_j$ values. The latter are
not much elongated along the $y$ axis. Due to this morphological
evolution of the x1 family there are no elliptical-like orbits
elongated along the $y$ axis to extend the bar towards corotation. The
elongated orbits which reach the farthest out in the $y$ direction are
elliptical-like orbits with loops, reaching $y \approx 9.4$, surrounded
by a roundish structure reaching the corotation region
(Fig.~\ref{a3model}).

\begin{figure}
\rotate[r]{
\epsfxsize=2.5cm \epsfbox{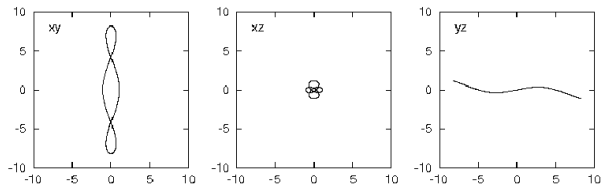}
}
\caption[]{Stable orbit of the x1v4 family of model A2.}
\label{A2ox1v4}
\end{figure}

Before closing our description of model A2, we should mention that the family
x1v4, initially bifurcated as {\em double
unstable}, becomes stable for larger energies and provides the system
with 3D orbits with low $\overline{|z|}$. The orbit we give in
Fig.~\ref{A2ox1v4} has $E_j =-0.12$,
while the family x1v4 bifurcates for $E_j \approx -0.245$ at a D\ar U
transition of x1. The evolution of the stability
indices of this family in model A2 is less complicated than in model
A1. It nevertheless shows all kinds of instabilities we encounter in
3D Hamiltonian systems and finally ends again {\em on} x1. This means
that it can be considered both as a direct and as an inverse bifurcation of
x1.

Summarizing the main differences of the orbital behaviour of the slow rotating
bar model from that in the fiducial case, we underline the existence of a
complicated common characteristic of the x1, x2 and x3 families. As a
consequence the simple-periodic families of the x1-tree appear in two parts.
The second part consists of x1$^{\prime}$ and its 3D bifurcations. The
families of the x1$^{\prime}$-tree have large stable parts, but they do not
help the bar reach closer to corotation since they are quasi-circular (or have
quasi-circular projections on the equatorial plane). The rectangular-like
orbits in this case are almost squares.  The model also includes a simple
periodic x2-like 3D family. Other differences in the orbital behaviour from
model A1 that should be mentioned is the small complex unstable part of x1v1
and the bifurcation of the family x1v4 at a D\ar U stability transition.

\subsection{A fast rotating bar}
Model A3 has a fast rotating bar. Its pattern speed is 0.0837, which
brings corotation to 4.2~kpc, i.e. closer to the center than the end of the
imposed bar. All other parameters remain as in models A1 and A2.

\begin{figure}
\rotate[r]{ 
\epsfxsize=6.1cm \epsfbox{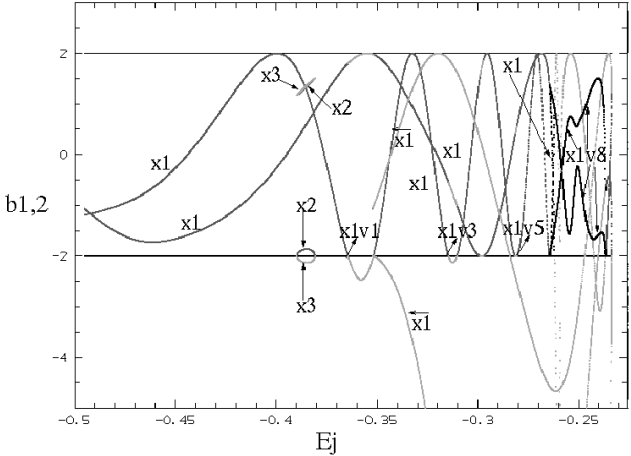}
}
\caption[]{Stability diagram for the family x1 in model A3. The black
bold curves at the right part of the figure ($-0.265<E_j<-0.235$) are
the stability indices of the family x1v8. Light grey curves indicate
instability.}
\label{A3stab}
\end{figure}

The major effect, as expected, is that the OILR approaches the IILR,
and the size of what we would call `x2-region' is drastically
reduced. In model A3 both x2 and x3 families still exist. The size of
the semi-major axis of the largest x2 orbit is 0.63~kpc. This means,
that the x2 orbits could support features of sizes about $\approx$
1.2~kpc. In other words in such models, the x2 family could play a
role in the dynamics of the innermost 1 kpc of the system if the corresponding
orbits are populated, despite the fact that the x2-loops we find are tiny 
($\Delta E_j \approx 0.01$) in
comparison to those of models A1 and of course A2.  In Fig.~\ref{A3stab} we
see the evolution of the stability indices of this model. Note the small
elliptical features around $E_j \approx -0.385$, which are made from the
combination of the stability indices of x2 and x3. The stability
indices of these two families do not have any other cuts or tangencies
with the $b=2$ or $b=-2$ axes and thus this model has no 3D families
oriented perpendicular to bar major axis and cannot form a peanut with
this orientation.   

The oscillations of the b1 and b2 curves of x1 bring in the system the
families x1v1, x1v3 and x1v5 as stable. Their dynamical behaviour, and
thus their importance for the global dynamics of the system, do not
differ from what we encountered in the fiducial case, and so we do not
discuss it further here. In this model x1v4 is not significant. It
remains unstable until its orbits reach high $z$ values above the
equatorial plane. The curves indicated by $\stackrel{\gets}{x1}$
correspond to the orbits at the branch of the characteristic of x1,
after the bend of the curve towards lower energies for $E_j \approx
-0.235$ (see Fig.~\ref{A3char} below). Light gray indicates also in
Fig.~\ref{A3stab} unstable orbits. The lower index almost goes through
the point of intersection of x1 with the $-2$ axis. However, because
of the location of the second index, we do not have a loop that closes
on x1 there.

The new elements that the study of this model brings to the
investigation of the orbital dynamics of barred potentials are
focused at the region of the (type 2) gap at the 4:1 resonance. In
Fig.~\ref{A3char} we show what is
\begin{figure}
\rotate[r]{
\epsfxsize=6.0cm \epsfbox{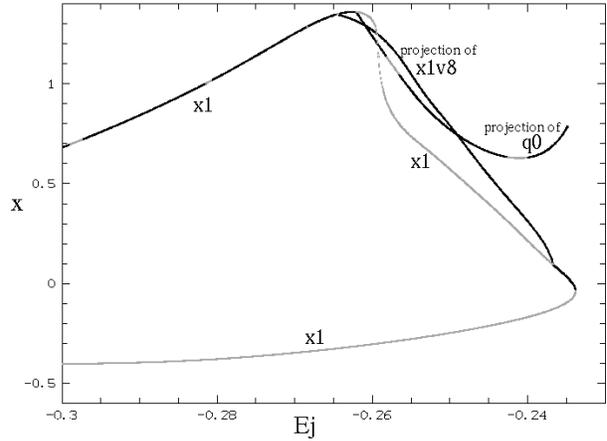}
}
\caption[]{Part of the characteristic diagram of model A3. It shows
the curve of the family x1 at the 4:1 resonance region, and the
($E_j$,$x$) projections of families q$_0$ and x1v8. Light
gray color indicates unstable orbits. }
\label{A3char}
\end{figure}
new in this model on a characteristic ($E_j$,$x$) diagram of x1. We
have also included the ($E_j$,$x$) projections of a planar family
(q$_0$), which has $\dot{x_0} \neq 0$ in the initial conditions, and a
3D family (x1v8), which is unstable in model A1. 

Let us start from the latter. As can be seen in Fig.~\ref{A3stab}, the
stability 
index associated with the vertical bifurcations has its seventh cut
with the $b=-2$ axis at $E_j\approx -0.265$ (the depth and size of the
unstable region is very small; we observe in Fig.~\ref{A3stab} that
the depth and size of the successive unstable regions decreases with
increasing energy).  At this point a new stable family is born.
Fig.~\ref{A3char} shows that this family is bifurcated just before the
local maximum of the x1 characteristic curve.  The $(x,z)$ and $(y,z)$
morphology of this new family is similar to that of family x1v8 of our
fiducial model (cf. Fig.~17c in paper I) and thus, according to the
rules set in paper I, we call it x1v8, although it emerges at the
seventh vertical bifurcation. 
In model A3 the succession of appearance of the bifurcating families
associated with the vertical 5:1 resonance is reversed compared with the
families of the corresponding instability strip in the fiducial case. Now this
instability strip is located before the local maximum of the x1-characteristic
at $E_j \approx -0.26$ (Fig.~\ref{A3char}), while in model A1 it is located
beyond the corresponding local maximum. As discussed in paper I when the
evolution of the stability index of x1 associated with the vertical
bifurcations has successive cuts with the $b=-2$ axis giving rise to a S\ar
U\ar S sequence in its stability, a stable and an unstable family are
introduced in the system. In model A1 for all instability strips at the
vertical resonances before the local maximum of the characteristic curve, the
families introduced as stable at the S\ar U transition are bifurcations in
$z$, and the unstable ones, bifurcated at the U\ar S part, are bifurcations in
$\dot{z}$. The opposite is true for the 5:1 resonance instability strip
located beyond the local maximum. There we had a stable family bifurcated in
$\dot{z}$, which we called x1v7 and an unstable one bifurcated in $z$ we
called x1v8 (paper I). In the present model the corresponding instability
strip of the vertical 5:1 resonance is located {\em before} the local maximum
of the x1-characteristic for $E_j \approx -0.26$ (Fig.~\ref{A3char}) and the
family introduced as stable is a bifurcation in $z$. Since we keep the
nomenclature introduced in the fiducial model throughout this series of
papers, this is family x1v8 and the bifurcation in $\dot{z}$, unstable in the
present model, is x1v7.
\begin{figure}
\rotate[r]{  
\epsfxsize=8.0cm \epsfbox{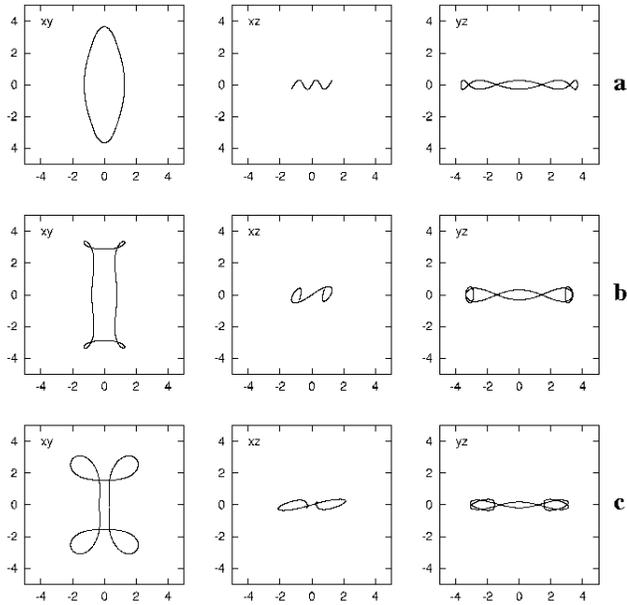}
}
\caption[]{Orbits of the family x1v8 of model A3. From top to bottom:
$E_j=-0.26$, $E_j=-0.25$ and $E_j=-0.24$ respectively.}
\label{ox1v7star}
\end{figure}
We note that while the x1v7 family of model A1 very soon gets orbits
with large $|z|$'s, family x1v8 is stable everywhere and its orbits
remain confined close to the equatorial plane (Fig.~\ref{ox1v7star}).

\begin{figure}
\rotate[r]{  
\epsfxsize=2.7cm \epsfbox{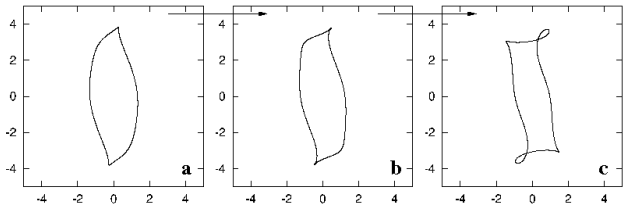}
}
\caption[]{A sequence of three stable orbits of the q$_0$ family of
model A3, showing its morphological evolution. They have (from left to
right) $E_j=-0.2615, -0.260$ and $-0.255$ respectively.}
\label{oq0}
\end{figure}

Almost at the local maximum of the x1 characteristic, at $E_j \approx
-0.26$, we have another S\ar U transition of x1
(Fig.~\ref{A3char}). There we have a radial bifurcation with ${\dot x}
\neq 0$ . We call the resulting family q$_0$, since it bifurcates at
the 4:1 resonance close to the local maximum, and its morphology is
different than that of the q1, q2 families of model A1. Its
morphological evolution, as we move along the ($E_j$,$x$) projection
of its characteristic, is given in Fig.~\ref{oq0}.
In Fig.~\ref{A3char} we see that q0 is stable almost everywhere with
only two small unstable zones. The one closer to the bifurcating point
is bridged by a family (not shown in Fig.~\ref{A3char}) existing just
in this interval.  The orbits of this family are slightly asymmetric
with respect to the corresponding unstable orbits of q0 at the same
$E_j$ values. Practically one could say that q0 is stable even
there. The second zone of instability is in an area where the loops of
q0 are big, so that the orbits are less interesting because of their
morphology. Thus we  conclude that practically all the
morphologically interesting parts of q0 are stable. 

Concluding about model A3, we can say that its differences with
respect to the fiducial case are focused at the dynamics close to the
local maximum of the characteristic at the radial 4:1 resonance. In
general, in most other models we examined, the decreasing branches of
the x1 characteristics beyond the local maximum harbor {\em mainly}
unstable families. In that respect, model A3 is an exception, because
at this region we can find decreasing branches with large stability
regions offered not by x1 itself but by q0 and x1v8.  We note that q0
and x1v8 are the most elongated rectangular-like orbits we found in
any of the models we studied. In model A3 the bar is supported by the x1 and
q0 families up to 3.8 kpc, with corotation at 4.2 kpc (For more details about
the supported face-on morphologies of our models, see Patsis, Skokos \&
Athanassoula, 2002b hereafter paper IV). Further differences are the size of
the x2 region, which in model A3 is very small and the insignificance of the
x1v4 family. A final difference of A3 from the rest of the models we examined
is the lack of a `bow' structure in the stability diagrams.  The rest of the
orbital structure is similar to that of model A1.

\section{A model without 2:1 resonances}
All models presented until now include an explicit bulge component in
the form of a Plummer sphere. In order to investigate the influence of
central concentrations on the dynamics of the bar we consider a model
without this component, and we increase the mass of the disc
accordingly so that the total mass stays the same as that of the other
models. This is model B, in which all other parameters are as in model
A1. We note that this particular case has been studied by Pfenniger (1984).

The model is characterized by the lack of radial as well as vertical
2:1 resonances. 
The stability indices of x1 have their first tangency with the $b=-2$ axis at
$E_j \approx -0.240$. We call the family bifurcated at this point  x1v5
because the $(x,z)$ and $(y,z)$ projections of its orbits have the same
morphology as that of the x1v5 family of model A1. x1v5 exists up to $E_j
\approx -0.219$  where it rejoins x1. This family corresponds to the B$z_1$
family of Pfenniger (1984). Another 3D orbit is bifurcated from x1 at $E_j
\approx -0.217$, and is morphologically similar to x1v5 so that we name it  
x1v5$^{\prime}$. This family has 
 stable orbits with low $|z|$ over a reasonable energy range, 
i.e it is an important family of
the system, as was already pointed out by Pfenniger (1984) who named it
 B$z_2$. At $E_j \approx -0.215$ the x1v7 family is bifurcated, which
corresponds to the B$\dot{z}_3$ family of  Pfenniger (1984). The overall
evolution of the stability indices in this model is characterized by a
complicated `bow', around $E_j \approx -0.22$, reminiscent of the bow
in model A1. The bow is at the center of the 3:1 region, which in this
model is rather extended. The values of the indices of the 3:1 families
remain smaller than 0, and all bifurcations are simple periodic
families. The model has t1, t2 and 3D 3:1 orbits with t1- and t2-like
projections. Its 4:1 gap is of type 2, and beyond this gap, towards
corotation, the orbital behaviour resembles that of model A1.

\begin{figure}
\rotate[r]{  
\epsfxsize=5.9cm \epsfbox{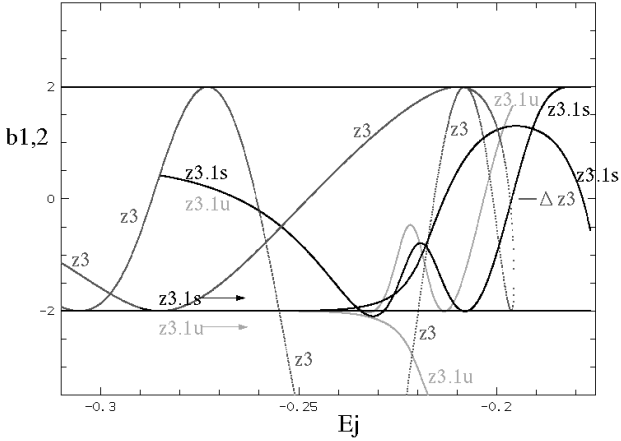}}
\caption[]{Part of the stability diagram of model B. It includes z3.1s,
  z3.1u and parts of z3.} 
\label{z3stab}
\end{figure}

In this model we found one more family, which has large stable parts
over a very extended energy range. This family has morphological
similarities with x1v4, but it is {\em not} related with the
x1-tree. This means that at least as far as we have followed, it does
not bifurcate from or be linked with  a family
belonging to the x1-tree. This family exists for $E_j >
-0.285$ and is one of the families of periodic orbits related to the z-axis
family, i.e. to the 1D orbits on the rotational axis of the system.  The well
known bifurcations of the z-axis family are the sao and uao families (Heisler,
Merritt \& Schwarzschild 1982). They are introduced at S\ar U and U\ar S
stability transitions of the z-axis family respectively, at which we
have cuts of one of the stability indices with the $b=-2$ axis.  We
find them considering the $z=0$ plane as a surface of section. However,
if the orbits of a single periodic family are repeated $n$-times it
can be considered as $n$-periodic (paper I, \S 2.2). As explained in
Appendix I, specific values of the stability index
(Eq.~\ref{sec:eq:be}) determine the $E_j$ value at which an
$n$-periodic family will bifurcate. These $n$-periodic families are the so called
`deuxi\`{e}me genre' families of Poincar\'{e} (1899).  The family we
found to be important in this model is a bifurcation of the z-axis
family when we consider its orbits repeated three times, i.e. of z3,
according to the nomenclature we introduced in paper I. In this case
tangencies of the stability indices of z3 with the $b=-2$ axis will
bifurcate two 3-periodic families and the family we discuss here is
one of them.  The z-axis family does not change its stability at the
energies at which the new families are born.  Already by studying the evolution
of the stability indices of the z-axis family, we can find out from
Eq.~\ref{sec:eq:be} the $E_j$ values at which 3-periodic bifurcations
will appear. Thus we know that for $E_j = -0.285$, a bifurcation of
the z-axis orbits with multiplicity 3 will be born.  We call this
family z3.1s and its position of birth is seen in Fig.~\ref{z3stab}.

In Fig.~\ref{z3stab} we give  the evolution of
the stability indices of z3. As expected at $E_j \approx -0.285$ it
has a tangency with the $b=-2$ axis, and z3.1s is bifurcated.
Actually at this point two families are bifurcated. At $E_j \approx
-0.256$ the one initially bifurcated as stable becomes unstable and
remains so thereafter, while the opposite is true for the one
bifurcated as unstable. We call z3.1s the one which is stable for the
largest energy range and z3.1u the one initially bifurcated as
stable. In any case their morphologies are very similar.  We note that
neither of the families have one of the stability indices on the $b=-2$
axis for some energy interval. Both stay close to it after the
bifurcating point, but not on it, as one can realize by looking at the
appropriate enlargement of Fig.~\ref{z3stab} (not plotted
here).  The range of energies over which z3.1s is stable emphasizes
its importance (Fig.~\ref{z3stab}).

\begin{figure}
\rotate[r]{  
\epsfxsize=7.5cm \epsfbox{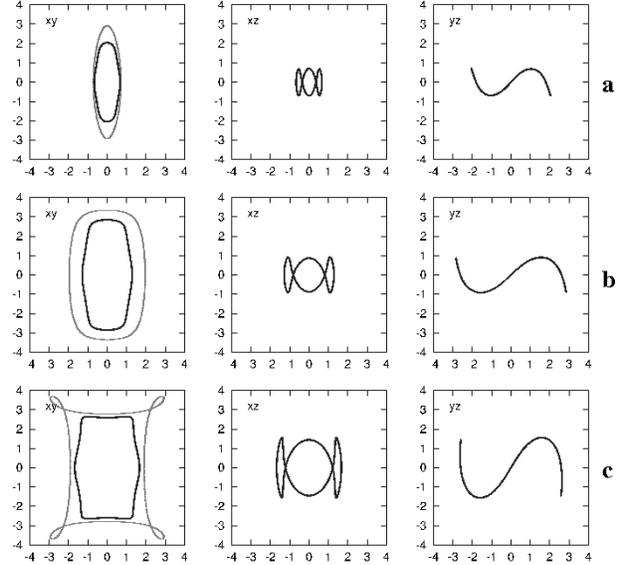}}
\caption[]{The morphological evolution of three stable z3.1s
  orbits. In (a) for $E_j = -0.25$, in (b) for $E_j = -0.22$, and in (c)
  for $E_j = -0.2$. In the $(x,y)$ projection we include, in grey, the
  x1 orbit with the same $E_j$.}
\label{oz3.1s}
\end{figure}

The multiplicity of an orbit is associated with the surface of section we
use. The z-axis orbits are calculated using as surface of section the
$(x,y)$ plane, so the multiplicity in this case does not refer to the
{\em morphology} of the projection of the orbit on this plane, but to
the number of intersections with this plane.  The detailed
morphological evolution is given in Fig.~\ref{oz3.1s}.  We observe
that the multiplicity of the z3.1s family if we consider as surface of
section the $y=0$ plane, as we do for all families of the x1-tree, is 1.
In the $(x,y)$ projection we have overplotted with light grey the
corresponding x1 orbits. The $(x,y)$ projection of z3.1s is always
included inside the curve of the x1 orbit. It is evident that the
morphology of the x1 orbits is similar but not identical to that of
the z3.1s $(x,y)$ projection. We observe also that the $(x,z)$ and $(y,z)$
projections remain close to the plane of symmetry of the galactic
model, at least for the lowest energies.

\begin{figure}
\rotate[r]{     
\epsfxsize=2.8cm \epsfbox{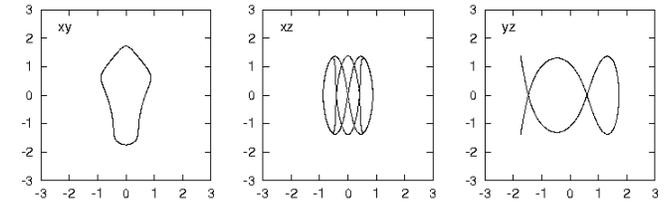}}
\caption[]{A stable z5.1s orbit at $E_j \approx -0.225$.} 
\label{oz5.1}
\end{figure}

Apart from z3.1s, we have found other families associated with z$n$s. A
case that could be mentioned is a bifurcation of z5, the shape of
which is given in Fig.~\ref{oz5.1}. 
Orbits like this can populate a galactic bulge or the central part of disks. Indeed,
although we have not an explicit bulge component in this model, our
disc is not flat. Due to the geometry of the Miyamoto disc one would
need in the central part orbits with projections on the $z$ axis of the
order of 1~kpc in order to build a self-consistent model. Thus orbits
like z5.1s should be considered.  In general, however, the tangencies
with the $b=-2$ axis are for larger energies and as a result, these
orbits, since they are bifurcations of the z-axis, have big
$\overline{|z|}$ values, and so, even if they have stable parts, are
not interesting building blocks for the disc of our
system\footnote{This is also the case for the z$n$ orbits for $n < 6$ in
all other models studied in this paper. Either they do not exist (models A1,
A2, A3 and D) or they are not so important because of their stability in
combination with their morphological evolution (model C).}.

For this model we underline the presence and importance of the z3.1s
orbits and the lack of the 3D families associated with low order
vertical resonances, since the first vertical bifurcation of x1 is
x1v5, a family bifurcated at the vertical 4:1 resonance.

\section{A model without radial ILRs}

Model C is intermediate between models A1 and B. It has a Plummer
sphere bulge, the scale length of which is 2.5 times larger than the
scale length of the bulge of A1. It is thus considerably less
centrally concentrated, and as a result its \O -$\kappa$/2 curve is
less peaked. This model does not have any radial ILRs, since we have,
like in model A1, \O$_b$=0.054.

\begin{figure}
\rotate[r]{
\epsfxsize=6.0cm \epsfbox{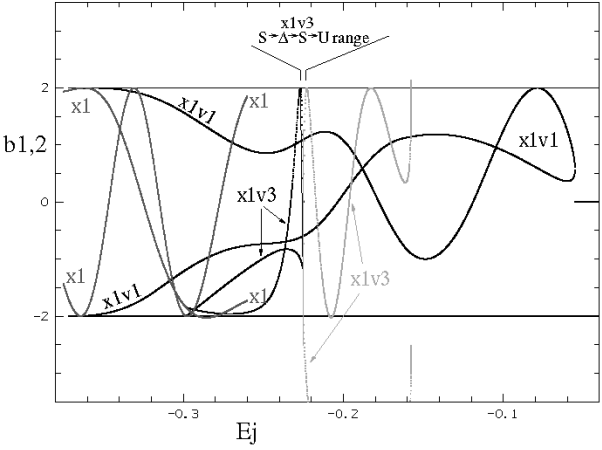}
}
\caption[]{Stability diagram for x1, x1v1 and x1v3 in model C.} 
\label{Bstab}
\end{figure}

\begin{figure}
\rotate[r]{   
\epsfxsize=2.8cm \epsfbox{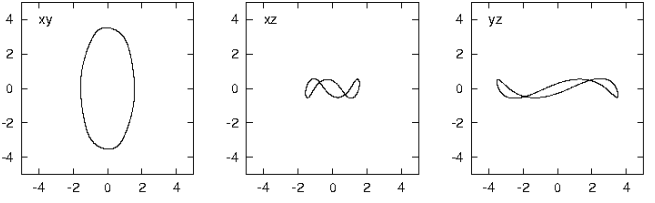}}
\caption[]{ A stable orbit of the family x1v3.1 in model C at $E_j
\approx -0.2235$.}  
\label{Bstrabh}
\end{figure}

On the other hand the model does have a vertical 2:1 resonance, 
where  is bifurcated a x1v1 family (Fig.~\ref{Bstab}),
which in this model is characterized by a large stable part.  After
the usual S\ar \D transition at $E_j = -0.08$ the family remains
always complex unstable. Furthermore, at $E_j = -0.26$ the maximum $z$
of the orbits is 1~kpc, and at $E_j$ = -0.225 the maximum $z$ is
1.5~kpc. This means that it is a very important family for the
dynamics of the system. x1v3 exists as well. It has a S\ar \Dd\ar S\ar
U sequence of stability types, but the \Dd\ar S\ar U part happens in a
very narrow energy range (Fig.~\ref{Bstab}). At the final S\ar U
transition the 3D family x1v3.1 depicted in Fig.~\ref{Bstrabh} is
bifurcated.
In this particular model this family is just bridging x1v3 with x1v4
at $E_j \approx -0.2232$. At this energy x1v4 becomes stable and
x1v3.1 can be considered as an inverse bifurcation of it. All this is
worth mentioning because x1v3.1 is a 3D family with a $(x,y)$ projection
resembling that of the family q0 of model A3.

The most important feature of model C is that x1v1 becomes complex
unstable for the first time at large energies and not just after it is
born as e.g. in the fiducial model A1. The consequences of this
stability evolution for the global dynamics of the model are described
in detail in paper III.

\section{A strong bar case}

\begin{figure}
\rotate[r]{        
\epsfxsize=6.0cm \epsfbox{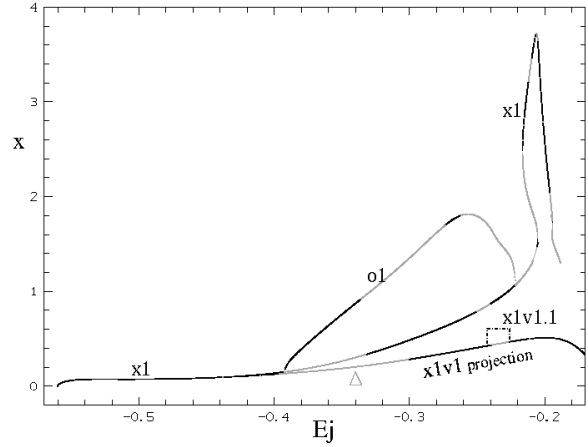}}
\caption[]{ Characteristic diagram for some important families in model D.
The unstable region of x1v1 bridged by x1v1.1 is indicated by a dashed-dotted
line. }  
\label{hdchar}
\end{figure}

Strong perturbations in Hamiltonian systems result in systems with a
larger degree of orbital instabilities, and a larger amount of chaos.
Model D has a bar twice as massive as that of the other models and a
disc accordingly less massive, so that the total mass is the same. We
can see the effect of this change in Fig.~\ref{hdchar}, which
is a characteristic diagram of families x1, o1 and also of the
($E_j$,$x$) projection of the characteristic of the 3D family
x1v1. The rising part of the branch of the x1 characteristic, for $E_j
<-0.205$, is steeper than in model A1. In this model x1 is mainly
stable at its decreasing branch ($E_j >-0.205$). The morphology of the
x1 orbits there is rectangular-like, and this clearly shows that the model
with a stronger bar favours this morphology.

\begin{figure}
\rotate[r]{ 
\epsfxsize=6.0cm \epsfbox{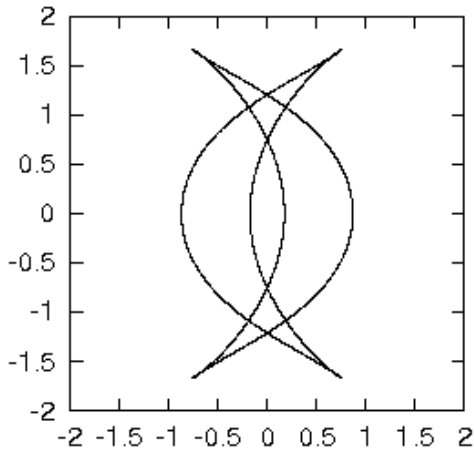}}
\caption[]{Two o1 orbits in model D, symmetric with respect to the
  major axis of the bar.  }
\label{t2hd}
\end{figure}

The behaviour of x2 in model D is similar to that in model A1. The variation
of the stability indices of x1 introduces as first bifurcating family in the
system family x1v1. This has first a short stable part and then becomes
complex unstable. The branch on Fig.~\ref{hdchar} indicated by x1v1 is just
the ($E_j$,$x$) projection of its characteristic curve. On this curve we note
with \D the complex unstable part. In the S\ar \D transition there is no
family inheriting the stability of x1v1 when the latter becomes unstable. As a
result, the only stable family for $-0.38 <E_j<-0.338$ is the o1 family, which
we also found in the slow bar case.  If, at a given energy, we consider the
two representatives of this family which are symmetric with respect to the
major axis of the bar, we get the combined morphology shown in
Fig.~\ref{t2hd}.

\begin{figure}
\rotate[r]{
\epsfxsize=5.0cm \epsfbox{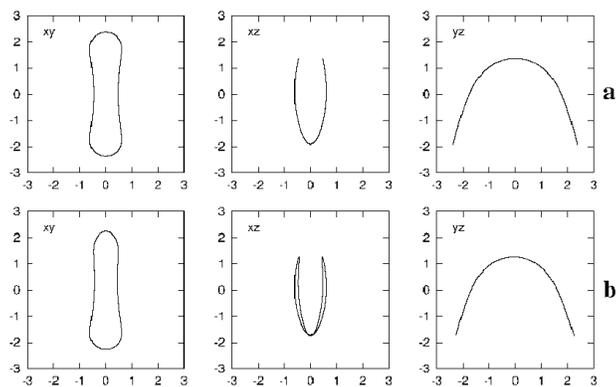}}
\caption[]{Orbits of the families x1v1 (a) and x1v1.1 (b) in model D
  at $ E_j \approx -0.225$ and $-0.235$ respectively.}
\label{ox1v1hd}
\end{figure}

We should also note that the $(x,y)$ projection of the x1v1 family, away from
the family's bifurcating point, does not quite follow the morphological
evolution of x1 at the same energy. The $(x,y)$ projections get squeezed on
the sides already before they become rectangular shaped and thus tend to take
a shape like `8'. This happens just before $|z|$ reaches values larger than
2~kpc. This morphology as well as the morphology of family x1v1.1, which
bridges a small zone of simple instability of x1v1, can be seen in
Fig.~\ref{ox1v1hd}.

To summarise the specific features of the orbital structure of 
model D it is worth  underlining that, due to its stronger bar, the
x1v1 family bifurcates at lower energies than in the other models. 
After it bifurcates from x1 as stable it
has, as usual, a complex unstable part, but
beyond the \Dd \ar S transition the orbits have still low
$\overline{|z|}$. Another interesting feature is the stability of the
x1 family at the decreasing part of the characteristic.  Also, as in model B,
  families x1v5 and x1v5$^{\prime}$ exist and have stable representatives.
Finally we note that from the families bifurcated initially as unstable, only
x1v6 has away from the bifurcating point a small stability part.

\section{Conclusions}
In this paper we  investigated the orbital structure in a class of
models representing 3D galactic bars. The parameters we varied are the
pattern speed, the strength of the bar and parameters defining the
bulge component of the galactic model. We found all the families that
could play an important role in the dynamics of 3D bars, and we
registered the main changes which happen as we vary the parameters
under consideration. Since evolutionary scenarios of the morphology of
the bars within a Hubble time could include an increase of the bulge
mass and a deceleration of the bar, as well as an increase or decrease
of the bar's strength, our models could correspond to discrete phases
in the dynamical evolution of a barred galaxy. They could thus be used
to explain the changes in the underling dynamics when the galaxies
evolve. Similarly, they can be used to understand the dynamics of
selected snapshots of $N$-body simulations.

Our main conclusions in the present paper are:
\begin{enumerate}
\item In all models we examined, the extent of the orbits which are
  most appropriate to sustain 3D bars is confined inside the radial
  4:1 resonance. Viewing the models face-on, the orbits with the
  longest projections along the major axis of the bar are either boxy
  or elongated with loops at the major axis; these are typical shapes
  of the orbits in the radial 4:1 resonance region. This behaviour is
  common to both slow and fast rotating bars.

\item The evolution of the characteristic of the basic family x1
  depends heavily on the pattern speed. The slower the bar rotates,
  the more complicated the x1-characteristic curve becomes. In the
  slowest of our models the families x1, x2 and x3 share the same
  characteristic curve. The folding of the characteristic towards
  lower energies, with most extreme case the one with the slow
  rotating bar, corresponds to a `bow' feature in the evolution of the
  stability indices as function of $E_j$. 

\item The fast rotating bar model A3 did not have the complicated
  evolution in the x1 characteristic and in the stability diagram
  corresponding to the `bow'. In this case all
  main 3D families of the x1-tree bifurcate from x1 at an S\ar U
  transition before the local maximum of the x1-characteristic at the
  radial 4:1 resonance and have initial conditions
  $(x,z,0,0)$.

\item The bars can be supported not only by x1-originated families
  but, depending on the model, by 3D orbits bifurcated from families
  related with the z-axis orbits. This has been encountered in the
  case of a model without radial or vertical 2:1 resonances.

\item Slow pattern rotation favours the presence of 3D x2-type orbits
  along the minor axis of the main bar. These orbits, which can lead to
  a 3D inner bar, are typical orbits of the potentials we studied.

\item The most elongated 4:1 rectangular-like orbits have been
  encountered in the fast rotating bar model A3. On the contrary, the
  corresponding orbits in the slow bar of model A2 are square-like and
  farther out circles and orbits with circular-like $(x,y)$
  projections. Thus in the slow bar case the bar is supported only by
  elliptical-like orbits of the x1-tree. The different
  elongations of the rectangular-like orbits can be explained by the
  fact that we have, in all models considered, bars of the same length
  in the imposed potential. Since the corotation radius changes with
  pattern speed, the non-axisymmetric part of the forcing is
  relatively larger near corotation for the fast bar than for the slow
  one.

\item The decreasing part of the x1-characteristic is in most cases
  unstable, except for the strong bar case (model D). This favours the
  presence of rectangular-like orbits at the outer parts of strong
  bars, in good agreement with observations (Athanassoula, Morin,
  Wozniak et al.  1990). This could be due to the fact that the bar
  forcing is stronger in the strong bar case. Stable rectangular-like
  orbits can also be found in the case of the fast rotating model A3,
  where rectangular-like stable orbits are provided not by the family
  x1 but by the families q0 and x1v8. This could again be due to the
  fact that, in the fast bar case, the forcing in the corotation
  region is larger than in other cases. The two above put together
  seem to argue that a strong bar forcing in the region around
  corotation is necessary for the model to have stable
  rectangular-like orbits which are sufficiently elongated along the
  bar major axis.

\item Models with low mass concentrations at the center (models B and
  C) favour the presence of $zn$ bifurcations for low $n$, which in
  some cases may be important for the global dynamics of the system
  (e.g. family z3.1s in model B). In this way we can have bar
  supporting families unrelated with the x1-tree.

\item In the x1-tree we encounter complex instability mainly in the
  x1v1 family. It can however, happen (e.g. in model C) that  complex
  instability appears in large energies and thus all orbits with
  $\overline{|z|} < 2$~kpc are stable. One must examine in every case
  the extent in $z$ of the complex unstable orbits of x1v1 in order to
  decide about the significance of this family for a model.

\end{enumerate}

The connection between the families of periodic orbits and the observed
morphologies in edge-on disc galaxies is discussed in paper III, and
the contribution of orbital theory to the question of the boxiness of
the outer isophotes in early type bars in paper IV.

\section*{Acknowledgments}
We acknowledge fruitful discussions and very useful comments by
Prof.~G.~Contopoulos. 
We thank the referee for useful suggestions that allowed to improve the
presentation of our work. This work has
been supported by E$\Pi$ET II and K$\Pi\Sigma$ 1994-1999;
and by the Research Committee of the Academy of Athens. Ch.~Skokos and
P.A.~Patsis thank the Laboratoire d'Astrophysique de Marseille, for an
invitation during which essential parts of this work have been completed.

\appendix

\section[]{ Poincar\'{e}'s `deuxi\`{e}me genre' families in
Hamiltonian systems}
The number \( n \) of intersections of a periodic orbit with the Poincar\'{e}
surface of section, when the orbit has a particular direction, defines its
multiplicity. So a periodic orbit of multiplicity \( n
\) has \( n \) points of intersection with the Poincar\'{e} surface of section
and it is  called a periodic orbit of period \( n \).

The linear stability or instability of a periodic orbit is defined by 
the eigenvalues of the corresponding monodromy matrix (see for example
Yakubovich \& Starzhinskii 1975).  The columns of this  matrix  are
suitably chosen   linearly independent solutions of the so-called variational
equations. These equations describe  the time evolution of a small deviation
from the periodic orbit. The eigenvalues of the monodromy matrix of a periodic
orbit can be grouped as pairs of inverse numbers, i.e. if \( \lambda \) is an
eigenvalue then \( 1/\lambda \) is also an eigenvalue (Broucke 1969,
Contopoulos \& Magnenat 1985).
The stability index \( b \) that corresponds to a particular pair of
eigenvalues is defined as:

\begin{equation}
\label{sec:eq:b}
b=-\left( \lambda +\frac{1}{\lambda }\right) 
\end{equation}
An orbit is stable when both stability indices are real numbers in the interval
\( (-2,\, 2) \), which equivalently means that the corresponding eigenvalues
are complex conjugate numbers on the unit circle.

As a parameter of the dynamical system changes the eigenvalues move on the
complex plane. When two eigenvalues moving on the unit circle coincide on \(
\lambda =1 \) and split along the real axis, the stability type of the orbit
changes from stable to unstable. The corresponding stability index is negative
and decreases below $-2$. At the same time a new periodic
orbit of the same multiplicity is born.
If, on the other hand, the two eigenvalues
continue to lie on the unit circle, after coinciding on \( \lambda =1 \),
which means that the orbit remains stable, then two new orbits of the same
multiplicity are born. 
A periodic orbit of multiplicity \( 1 \) can be also considered as a periodic
orbit of multiplicity \( n>1 \) if it is repeated $n$ times. It has as
monodromy matrix \( M_{n} \) the matrix
{\begin{figure}
\epsfxsize=4.0cm \epsfbox{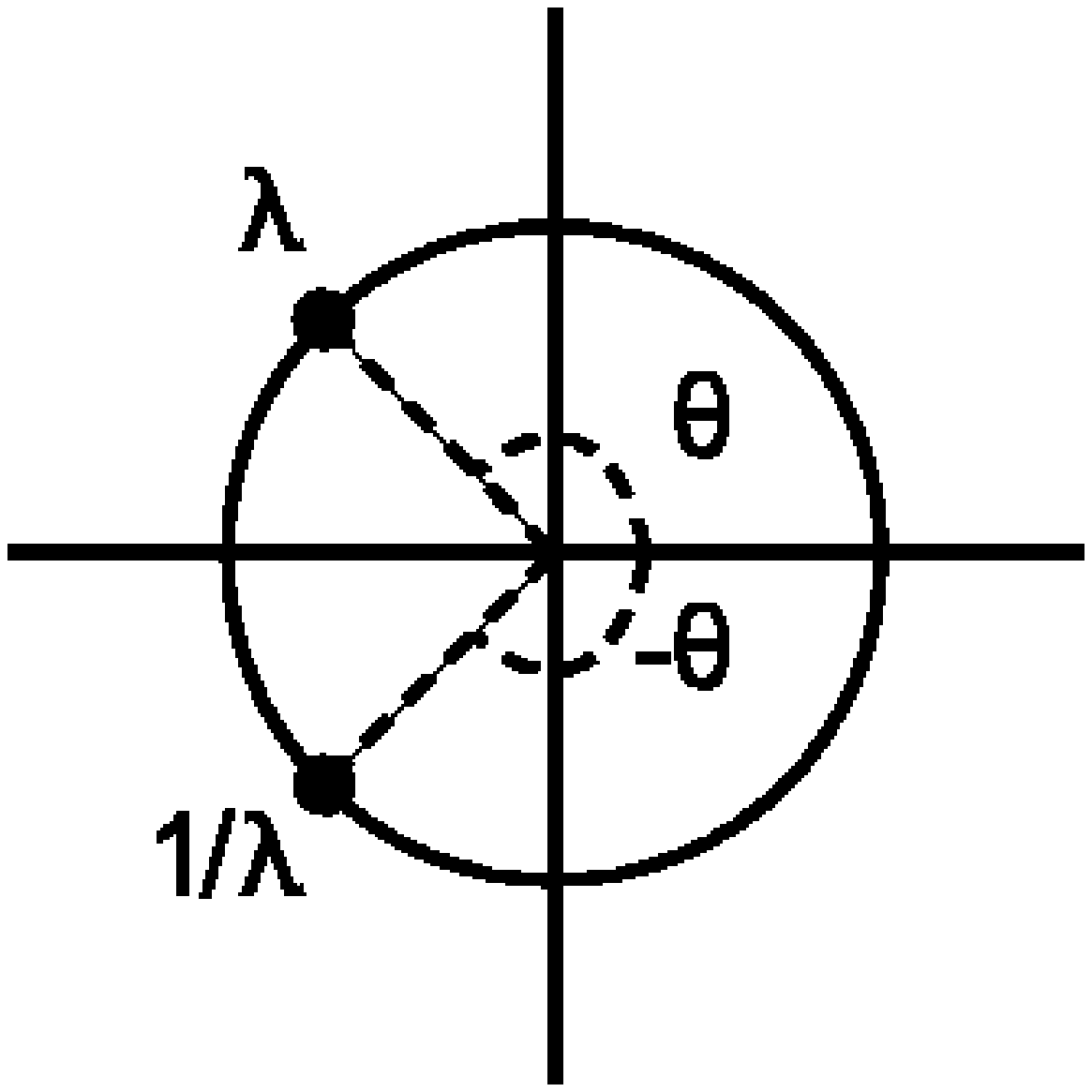}
\caption{\label{fig: eigenvalues}The eigenvalues of a stable orbit on
the unit circle.}
\end{figure}\par}
\begin{equation}
\label{sec:eq:A}
M_{n}=M^{n}_{1},
\end{equation}
where \( M_{1} \) is the monodromy matrix of the periodic orbit considered as
1-periodic. If the periodic orbit of period \( 1 \) is stable, then it has a pair
of eigenvalues of the form 
\begin{equation}
\label{sec:eq:lambda}
\lambda =\cos \vartheta +i\sin \vartheta \, ,\: \: \: \frac{1}{\lambda }=\cos
\vartheta -i\sin \vartheta \, , 
\end{equation}
as seen in Fig. \ref{fig: eigenvalues}. Thus the corresponding stability
index is:
\begin{equation}
\label{sec:eq:bb}
b=-\left( \lambda +\frac{1}{\lambda }\right) =-2\cos \vartheta .
\end{equation}
Considering this orbit as one of period \( n \) its eigenvalues will be of the
form \( \lambda ^{n} \), \( (1/\lambda )^{n} \), so that the corresponding
stability index \( b_{(n)} \)becomes:
\begin{equation}
\label{sec:eq:bn}
b_{(n)}=-2\cos (n\vartheta ).
\end{equation}
A tangency of \( b_{(n)} \) with the line \( b=-2 \) gives birth to two
new periodic orbits of period \( n \), while the 1-periodic orbit
remains stable. This bifurcation happens when 

\begin{equation}
\label{sec:eq:theta}
b_{(n)}=-2\Rightarrow \cos (n\vartheta )=1
\end{equation}
This condition is satisfied if we have e.g. 
\begin{equation}
\vartheta =2\pi
\frac{1}{n}, 
\end{equation}                                          
or equivalently when the stability index \( b \) of the period \( 1 \)
periodic 
orbit crosses the line
\begin{equation}
\label{sec:eq:be}
b=-2\cos \left( 2\pi \frac{1}{n}\right) .
\end{equation}
\begin{figure}
\epsfxsize=6.0cm \epsfbox{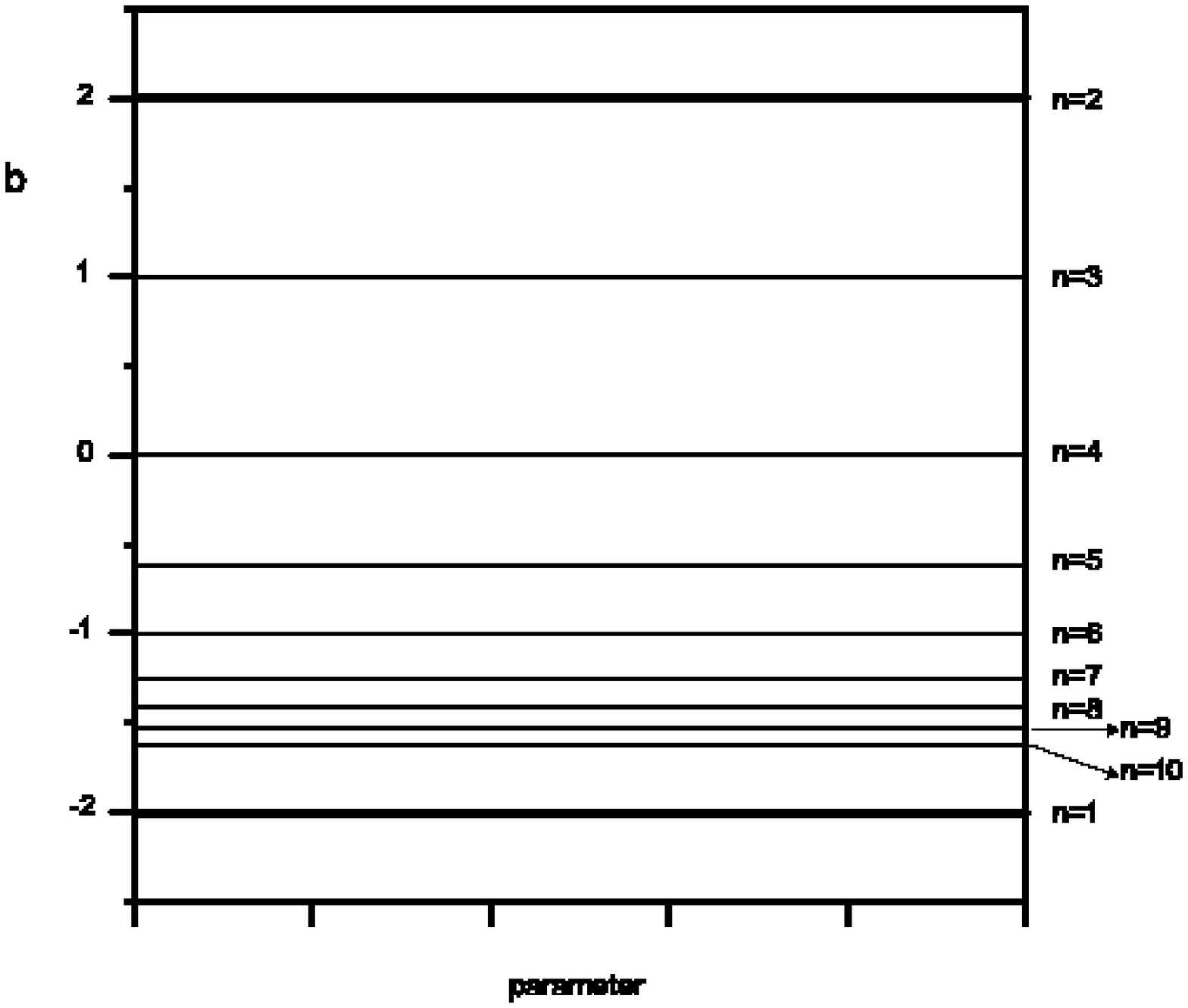}
\caption{\label{fig: lines}The values of the stability index \protect\(
  b\protect \)
that correspond to bifurcations of period \protect\( n=1,2,...,10\protect \)
given by Eq. (\ref{sec:eq:be}).}
\end{figure}
The bifurcating families of periodic orbits are the 
`deuxi\`{e}me genre' families of Poincar\'{e} (1899).  In Fig. \ref{fig: lines}
we plot the lines given by (\ref{sec:eq:be}) corresponding for \( n =
1,2,... 10 \). We see that, as we approach $b=-2$, the density of the lines  
 as well as the period of the bifurcating orbit increase.

\bsp

\label{lastpage}

\end{document}